\documentclass[journal]{IEEEtran}

\usepackage{color}
\usepackage{cite}
\newcommand{\triblue}{\color{blue}$\boldsymbol{\bigtriangleup}$\color{black}}
\newcommand{\trired}{\color{red}$\boldsymbol{\bigtriangledown}$\color{black}}

\newcommand{\tripur}{\color{mypur}\Large$\boldsymbol{\lhd}$\color{black}\normalsize}
\definecolor{mygreen}{rgb}{0, 0.4, 0}
\definecolor{mypur}{rgb}{0.4, 0, 0.4}

\ifCLASSINFOpdf
 \usepackage[pdftex]{graphicx}
\else
\fi
\usepackage{amssymb}
\usepackage{amsmath}
\usepackage{array}
\usepackage[caption=false,font=footnotesize]{subfig}
\usepackage{dblfloatfix}

\ifCLASSOPTIONcaptionsoff
  \usepackage[nomarkers]{endfloat}
 \let\MYoriglatexcaption\caption
 \renewcommand{\caption}[2][\relax]{\MYoriglatexcaption[#2]{#2}}
\fi
\usepackage{url}

\begin{document}

\title{Sparse Sensing with Semi-Coprime Arrays}

\author{Kaushallya~Adhikari

\thanks{This material is based upon research supported by the Louisiana Tech University.}        
\thanks{The author is with the  Louisiana Tech University, Ruston, LA, 71270 USA. e-mail: adhikari@latech.edu.}}

\maketitle

\begin{abstract}
A semi-coprime array (SCA) interleaves two undersampled uniform linear arrays (ULAs) and a $Q$ element standard ULA. The undersampling factors of the first two arrays are $QM$ and $QN$ respectively where $M$ and $N$ are coprime. The resulting non-uniform linear array is highly sparse. Taking the minimum of the absolute values of the conventional beampatterns of the three arrays results in a beampattern free of grating lobes. The SCA offers more savings in the number of sensors than other popular sparse arrays like coprime arrays, nested arrays, and minimum redundant arrays. Also, the SCA exhibits better side lobe patterns than other sparse arrays. An example of direction of arrival estimation with the SCA illustrates SCA's promising potential in reducing number of sensors, decreasing system cost and complexity in various signal sensing and processing applications.

\end{abstract}


\IEEEpeerreviewmaketitle

\section{Introduction}
\label{sec:intro}

\IEEEPARstart{M}{any} linear sparse array designs achieve the resolution of a fully populated uniform linear array (ULA), hereafter called full ULA  \cite{Kpt,Unz,Ishimaru,IshimaruChen,Moffet,steinberg,berman, davis, kefalas,mitra,nested1,nested2,VandP1,VandP2}. Some of these designs, \cite{Kpt,Unz,Ishimaru,IshimaruChen,Moffet}, use conventional beamforming (CBF) to process the received signal, while some other designs, \cite{steinberg,berman, davis, kefalas,mitra,nested1,nested2,VandP1,VandP2}, split the sparse array into two subarrays and multiply the subarrays' individual CBF outputs. The sparse array designs that existed prior to 2010 suffered from a common limitation --- the lack of concrete design criteria or the analytical expressions for the sensor locations.  The nested arrays \cite{nested1,nested2} and coprime sensor arrays (CSAs) \cite{VandP1,VandP2} overcome the limitation by providing analytical expressions for the sensor locations. The generalized coprime array configuration treats the coprime and nested arrays as its special cases \cite{QinZhangAmin}.

A nested array interleaves a short full ULA with an undersampled ULA having the same undersampling factor as the number of sensors in the short full ULA. A CSA interleaves two undersampled subarrays where the undersampling factors are coprime integers. In addition to having convenient analytical expressions for the sensor locations, the NSA and CSA also have a clear mechanism to disambiguate aliasing that occurs due to undersampling. Moreover, the CSAs can also match the peak side lobe height of a full ULA of equal resolution \cite{KBW} thereby dominating the field of sparse arrays in recent years. The reduction of number of sensors in these sparse arrays translates to lower system cost, and decreased system complexity.

This paper introduces a novel sparse array called semi-coprime array (SCA). The SCA has the potential to offer more reduction in the number of sensors than even the CSA and the NSA, while still offering the crucial advantages of the CSA and the NSA which are the analytical expressions for the sensor locations, equal resolution as the full ULA with equivalent aperture, and prudent cancellation of the grating lobes resulting from undersampling. The SCA can achieve the peak side lobe height of a full ULA with much less extension than the CSA and the NSA. Since the SCA is sparser than the CSA and the NSA, the SCA suffers less from mutual coupling effect between adjacent sensors than the CSA and the NSA.

Section \ref{sec:sca} describes the SCA in detail. Section \ref{sec:comparison} compares the SCA with other popular sparse arrays (CSA, NSA, and minimum redundant arrays).  Section \ref{sec:results} demonstrates how SCA's inherent super-resolution characteristic can be used in direction of arrival estimation and compares the SCA with other sparse arrays.

\textit{Conventions:} Bold-faced letters represent vectors; $x^*$ denotes complex conjugate of $x$; $\textbf{x}^H$ denotes Hermitian of $\textbf{x}$; $GCD(a,b)$ denotes the greatest common divisor (GCD) of the integers $a$ and $b$; $min(a,b,c)$ denotes the minimum of $a$, $b$, and $c$.

\section{Semi-Coprime Arrays}
\label{sec:sca}
 
This section is going to describe the novel array design and its associated processor in detail.

\subsection{Semi-Coprime Arrays Structure}

\begin{figure*}
\centering
\includegraphics[scale=0.32,trim = 0cm 0cm 0cm 0cm]{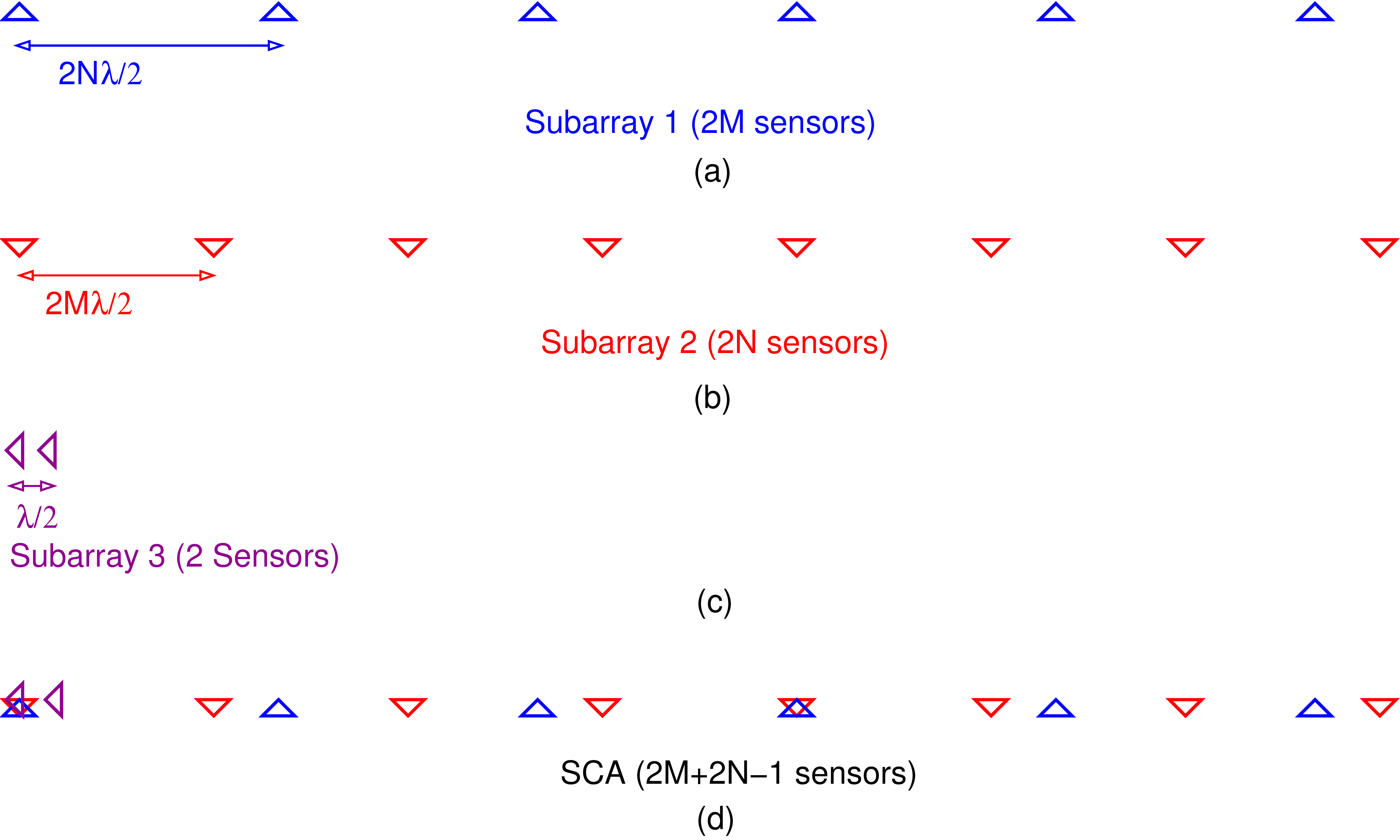}
\caption{(a). Subarray 1 with $2M$ sensors and undersampling factor $2N$ (b). Subarray 2 with $2N$ sensors and undersampling factor $2M$ (c). Subarray 3 with $2$ sensors and undersampling factor $1$ (d). The SCA resulting from interleaving Subarray 1, Subarray 2, and Subarray 3}
\label{scaarray}
\end{figure*}

A semi-coprime array (SCA) is a sparse array that interleaves three ULAs, hereafter called Subarray 1, Subarray 2, and Subarray 3. Each SCA has underlying coprime integers $M,$ and $N.$ Subarray 1 has $PM$ sensors (Symbol \color{blue}\triblue\color{black}) and $QN\dfrac{\lambda}{2}$ intersensor spacing, and Subarray 2 has $PN$ sensors (Symbol \color{red}\trired\color{black}) and $QM\dfrac{\lambda}{2}$ intersensor spacing, where $P$ and $Q$ are integers greater than $1$, and $\lambda$ is the wavelength of the signal to be sampled. The Subarray 3 has $\dfrac{\lambda}{2}$ intersensor spacing and the number of sensors (Symbol \color{mypur}\tripur\color{black}) is equal to the GCD of the undersampling factors in the Subarray 1 and Subarray 2, i.e., $Q=GCD(QM,QN).$ Figure \ref{scaarray} depicts the formation of an SCA for $M=3,$ $N=4,$ $P=2,$ and $Q=2.$ The three subarrays always share the first sensor. The Subarray 1  and the Subarray 2  always share $P$ sensors. Hence, the total number of sensors in an SCA is $PM+PN+Q-1-P$ but it can achieve the resolution of a full ULA with $PQMN$ sensors. For example, if $P=2$, $Q=6$, $M=4$, and $N=5$, the SCA has only $21$ sensors and it achieves the resolution of a full ULA with $240$ sensors.

\subsection{Min Processing}
\label{sec:minprocessing}

\begin{figure}[h]
\centering
\includegraphics[scale=0.32,trim = 0cm 0cm 0cm 0cm]{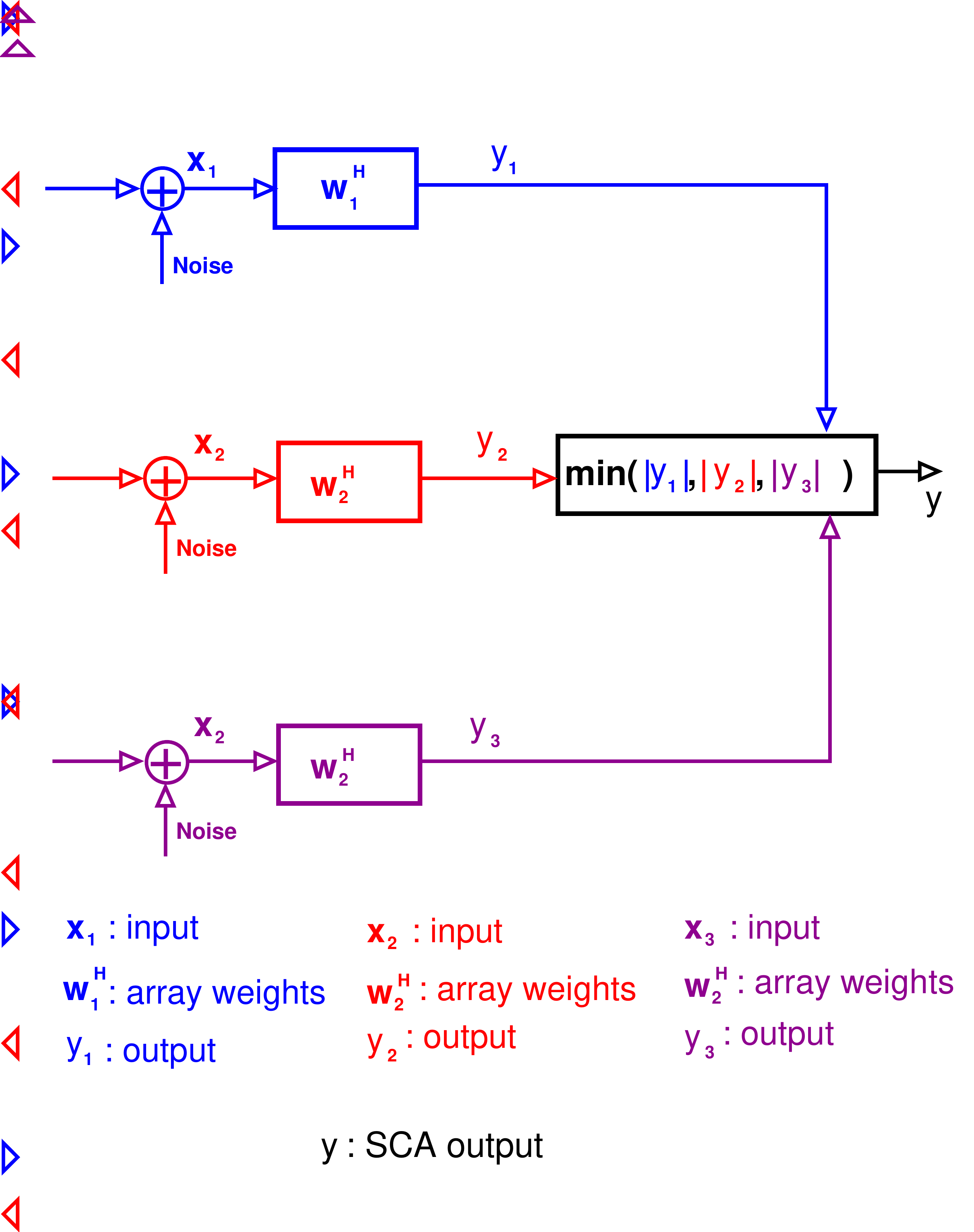}
\caption{The SCA min processor conventionally beamforms the Subarray 1 (blue), Subarray 2 (red) and Subarray 3 (purple) data and finds the minimum of the absolute values of the CBF outputs to produce the final output.}
\label{scaprocessor}
\end{figure}

The proposed processor for the SCA is a min processor as depicted in Figure \ref{scaprocessor}. The vectors $\textbf{x}_1,$ $\textbf{x}_2,$ and $\textbf{x}_3$ represent the signal received by the three subarrays. The SCA processor conventionally beamforms each of the three subarrays' received signal by using the weight vectors $\textbf{w}_1,$ $\textbf{w}_2,$ and $\textbf{w}_3$. Assuming uniform weighting, the vectors $\textbf{w}_1,$ $\textbf{w}_2,$ and $\textbf{w}_3$ are $PM$, $PN$, and $Q$ element vectors and their $i^{th}$ elements for direction cosine $u=\cos(\theta)$ are $\dfrac{1}{PM}\exp(j\pi u (i-1)QN)$, $\dfrac{1}{PN}\exp(j\pi u (i-1)QM)$, and $\dfrac{1}{Q}\exp(j\pi u (i-1))$ respectively. The CBF outputs for the three subarrays are $y_1=\textbf{w}_1^H\textbf{x}_1,$ $y_2=\textbf{w}_2^H\textbf{x}_2,$ and $y_3=\textbf{w}_3^H\textbf{x}_3$. The final SCA output, $y,$ is the minimum of the absolute values of the three CBF outputs, i.e., $y=min(|y_1|,|y_2|,|y_3|).$

Although both Subarray 1 and Subarray 2 are undersampled, the SCA output disambiguates aliasing by appropriately combining the Subarray 1, Subarray 2 and Subarray 3 outputs. The Subarray 1 beampattern has the undersampling factor $QN,$ and as a result, it has $QN$ major lobes at integer multiples of $2/(QN).$ Assuming the array is steered to broadside, i.e. $u=\cos\left(\dfrac{\pi}{2}\right)=0$, the major lobe at $u=0$ is the main lobe and the other $QN-1$ major lobes are grating lobes resulting from undersampling.  The Subarray 2 beampattern  has the undersampling factor $QM,$ and consequently, it has $QM$ major lobes at integer multiples of $2/(QM).$ The major lobe at $u=0$ is the main lobe and the other $QM-1$ major lobes are grating lobes due to undersampling. The Subarray 3 beampattern has one major lobe at $u=0$ which is the main lobe and it has nulls at integer multiples of $2/Q$. Since the undersampling factors of Subarray 1 and Subarray 2 have $GCD(QM,QN)=Q,$ only $Q$ major lobes of Subarray 1 overlap with $Q$ major lobes of Subarray 2. One of the overlapping major lobes from each of the Subarray 1 and Subarray 2 is the main lobe at $u=0,$ and the other $Q-1$ overlapping major lobes from each of the Subarray 1 and Subarray 2 cause aliasing. However, the Subarray 3 has nulls exactly at the locations where the other two subarrays have overlapping grating lobes. As a result, taking the minimum of the three beampatterns generates a beampattern with no grating lobes at all.

\begin{figure}[h]
\centering
\includegraphics[scale=0.4,trim = 0cm 4cm 0cm 4cm]{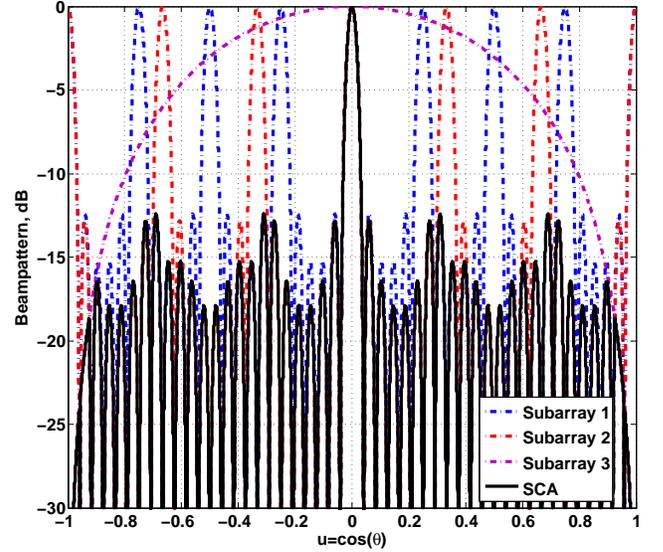}
\caption{Formation of the SCA beampattern. Taking the minimum of the absolute values of the Subarray 1 (blue dashed-dot), Subarray 2 (red dashed-dot) and Subarray 3 (purple dashed-dot) beampatterns results in a beampattern (black solid) free of grating lobes.}
\label{scabeampattern1}
\end{figure}

\begin{figure}[h]
\centering
\includegraphics[scale=0.4,trim = 0cm 4cm 0cm 4cm]{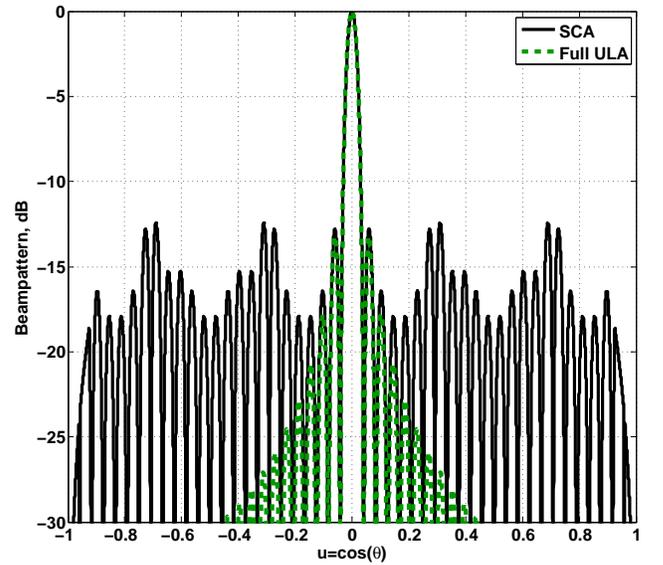}
\caption{Comparison of the SCA (black solid), and ULA (green dashed) beampatterns. The two beampatterns have equal main lobe with and PSL height.}
\label{scabeampattern2}
\end{figure}

Figure \ref{scabeampattern1} elucidates the beampattern formation mechanism for the SCA shown in Figure \ref{scaarray}, assuming the array is steered to the direction $u=0.$ The Subarray 1 beampattern (blue dashed in Figure \ref{scabeampattern1}) has the undersampling factor $2N=8,$ and as a result, it has $8$ major lobes. The major lobe locations are integer multiples of $1/N,$ i.e., $u=-0.75,$ $u=-0.5,$ $u=-0.25,$ $u=0,$ $u=0.25,$ $u=0.5,$ $u=0.75,$ and $u=\pm{1}.$ The major lobe at $u=0$ is the main lobe and the other $7$ major lobes are grating lobes resulting from undersampling.  The Subarray 2 beampattern (red dashed-dot  in Figure \ref{scabeampattern1}) has the undersampling factor $2M=6,$ and consequently, it has $6$ major lobes. The major lobe locations are integer multiples of $1/M,$ i.e., $u=-0.67,$ $u=-0.33,$ $u=0,$ $u=0.33,$ $u=0.67,$ and $u=\pm{1}.$ The major lobe at $u=0$ is the main lobe and the other $5$ major lobes are grating lobes due to undersampling. The Subarray 3 beampattern (purple dot in Figure \ref{scabeampattern1}) has one major lobe at $u=0$ which is the main lobe. Since the undersampling factors of Subarray 1 and Subarray 2 have $GCD(QM,QN)=2,$ two major lobes of Subarray 1 overlap with two major lobes of Subarray 2. One of the overlapping major lobes from each of the Subarray 1 and Subarray 2 is the main lobe at $u=0,$ and the other overlapping major lobe from each of the Subarray 1 and Subarray 2 is the grating lobe at $u=\pm{1}.$ However, the Subarray 3 has nulls exactly at $u=\pm{1}$. Therefore, taking the minimum of the three beampatterns shown in Figure \ref{scabeampattern1} results in a beampattern (black solid in Figure \ref{scabeampattern1}) that is devoid of grating lobes.

Figure \ref{scabeampattern2} compares an SCA beampattern with a standard ULA that offers the same resolution. The standard ULA (green dashed in Figure \ref{scabeampattern2}) has $48$ sensors while the SCA (black solid in Figure \ref{scabeampattern2}) has only $13$ sensors. The SCA and the standard ULA have equal main lobe widths and hence equal resolution, and almost equal peak side lobe (PSL) heights. Using only about $27\%$ of the sensors in the standard ULA, the SCA is able to match both resolution and PSL height of the standard ULA, offering a substantial saving in the number of sensors.

\section{Comparison with Other Sparse Arrays}
\label{sec:comparison}

As noted in the introduction, many authors have proposed various sparse array schemes over the course of the last few decades. However, the introduction of the NSA and CSA in 2010 and 2011 has sparked renewed interest in sparse arrays. The CSA and NSA have drawn great attention from researchers since they possess concrete expressions for sensor locations and their exists a clear mechanism to remove prudently any grating lobes that arise due to undersampling.  Another category of arrays, minimum redundant arrays (MRAs), has also been a relevant topic of research and reference in the field of sparse arrays for several decades. This section summarizes three different versions of the CSA (basic CSA, extended CSA, and min-processing CSA), the NSA, and the MRA and compare them with the novel array, SCA.
  
\subsubsection{Basic Coprime Sensor Array}
\label{sec:bcsa}

\begin{figure}[!h]
\centering
\includegraphics[scale=0.3,trim = 0cm 0cm 0cm 0cm]{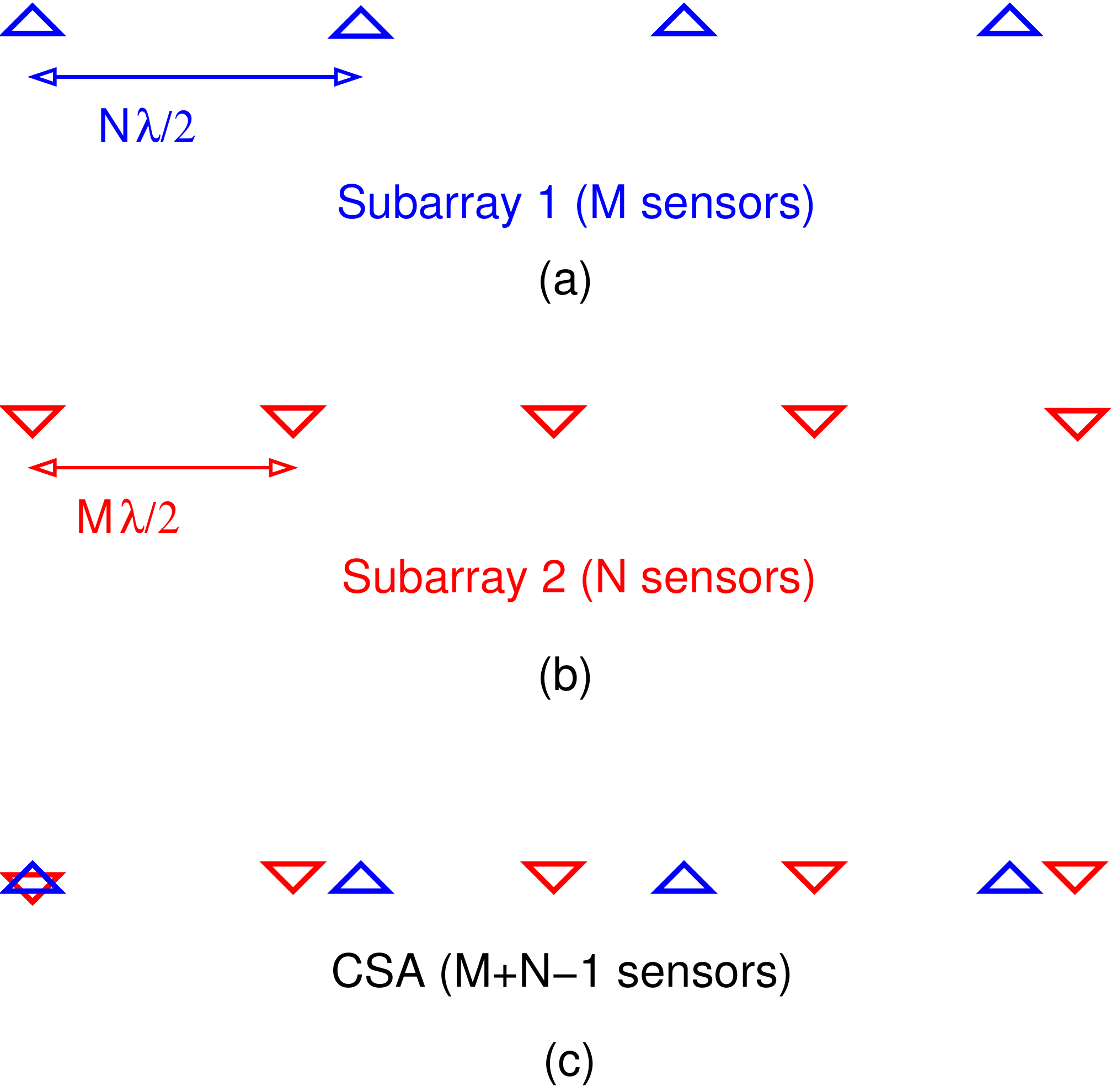}
\caption{(a) Subarray 1 (blue) of a CSA with coprime pair $(4,5)$ (2) Subarray 2 (red) of a CSA with coprime pair $(4,5)$ (c) The resulting CSA formed by interleaving Subarray 1 and Subarray 2.}
\label{csasensors}
\end{figure}

A basic CSA is a sparse array that interleaves two ULAs, hereafter called Subarray 1, and Subarray 2. Each CSA has underlying coprime integers $M,$ and $N.$ The Subarray 1 has $M$ sensors (Symbol \color{blue}\triblue \color{black} \hspace*{0.03cm} in Figure \ref{csasensors}) and $N\dfrac{\lambda}{2}$ intersensor spacing, and the Subarray 2 has $N$ sensors (Symbol \color{red}\trired \color{black} \hspace*{0.03cm} in Figure \ref{csasensors}) and $M\dfrac{\lambda}{2}$ intersensor spacing. The total number of sensors in a basic CSA is $M+N-1$ but it can achieve the resolution of a full ULA with $MN$ sensors \cite{VandP1,VandP2}. Figure \ref{csasensors} depicts the formation of a basic CSA for $M=4,$ and $N=5.$  The basic CSA has only $8$ sensors and it achieves the resolution of a full ULA with $20$ sensors.  For a CSA with a given aperture, the coprime pair $M$ and $N=M+1$ minimizes the total number of sensors needed to span that aperture \cite{KBW,shadings}. This paper assumes that $N=M+1$.

A CSA processor eliminates the grating lobes by taking the product of the CBF beampatterns of the two subarrays. The Subarray 1 beampattern has the undersampling factor $N,$ and as a result, it has $N$ major lobes at integer multiples of $2/N.$ The major lobe at $u=0$ is the main lobe and the other $N-1$ major lobes are grating lobes due to undersampling.  The Subarray 2 beampattern  has the undersampling factor $M,$ and as a result, it has $M$ major lobes at integer multiples of $2/M.$ The major lobe at $u=0$ is the main lobe and the other $M-1$ major lobes are grating lobes due to undersampling. Since $M$ and $N$ are coprime, all the grating lobe locations are unique while the main lobes of the Subarray 1 and Subarray 2 are exactly at the same location. Taking the product of the two beampatterns removes the grating lobes.

\begin{figure}[!h]
\centering
\includegraphics[scale=0.4,trim = 0cm 4.5cm 0cm 4.5cm]{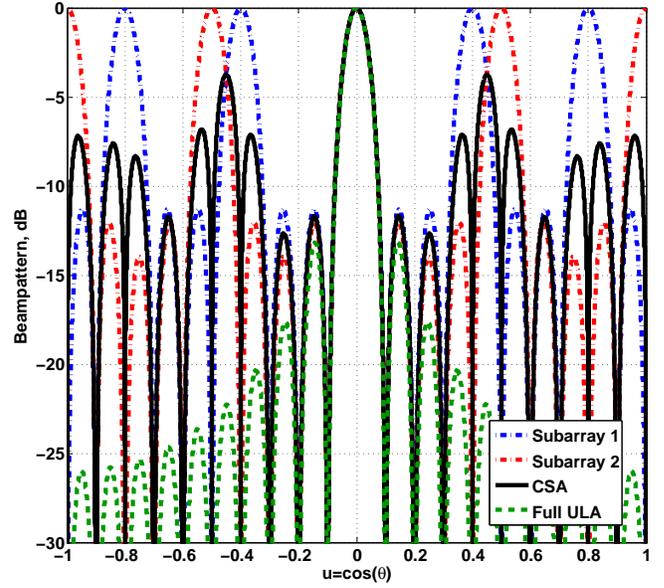}
\caption{Subarrays, CSA and full ULA beampatterns.}
\label{basiccsabp}
\end{figure}

Figure \ref{basiccsabp} illustrates the formation of a CSA beampattern for the arrays shown in Figure \ref{csasensors}. The CSA has total $8$ sensors and the ULA has $20$ sensors. However, they have the equal main lobe width and therefore, have equal resolution. Comparing the CSA and the ULA beampatterns also shows that the CSA peak side lobe (PSL) height is much higher than the ULA PSL height. The ULA PSL height is $-13$ \rm{dB} while the CSA PSL height is about $-4$ \rm{dB} in Figure \ref{basiccsabp}. 

\subsubsection{Extended Coprime Sensor Array}

As noted in Section \ref{sec:bcsa}, the PSL height of a basic CSA is too high for it to be considered useful in major applications. Keeping the intersensor spacings of the subarrays fixed and adding more sensors to both subarrays causes the CSA PSL to decrease and match the ULA PSL height \cite{VandP1,VandP2}. The number of additional sensors required to be added to the subarrays to match the PSL height of a standard ULA depends on the coprime factors and the weighting functions used. The derivation of the number of additional sensors for various standard uniform and non-uniform weighting functions exists in \cite{shadings}.  The coprime sensor array where the subarrays have been extended to match the PSL height of the full ULA with the equivalent aperture is called the extended coprime sensor array (ECSA). The numbers of sensors in the Subarray 1 and Subarray 2 are $M_e=\lceil cN\rceil -1$ and $N_e=\lceil cN\rceil$. The constant $c$ is called the extension factor which is the number of repetitions of the basic CSA in an ECSA. For uniform shading, the total number of sensors in an ECSA is $L=13M+6$ and the ECSA has the resolution of a full ULA with $M_eN$ sensors (See Appendix \ref{app:ecsa}).

\subsubsection{Min-processing Coprime Sensor Array}

Taking the minimum of the two CBF beampatterns in a basic CSA or an ECSA also removes the grating lobes since the grating lobes are at different locations \cite{liubuck1,liubuck2,liubuck3}. A CSA where each subarray has two periods of the basic subarrays and the associated processor is the min-processor explained in \cite{liubuck1,liubuck2,liubuck3} is, subsequently, called min-processing comprime sensor array (MCSA). An MCSA has $4M$ sensors and it achieves the resolution of a full ULA with $2MN$ sensors (See Appendix \ref{app:mcsa}).

\subsubsection{Nested Sensor Array}

\begin{figure}[!h]
\centering
\includegraphics[scale=0.3,trim = 0cm 0cm 0cm 0cm]{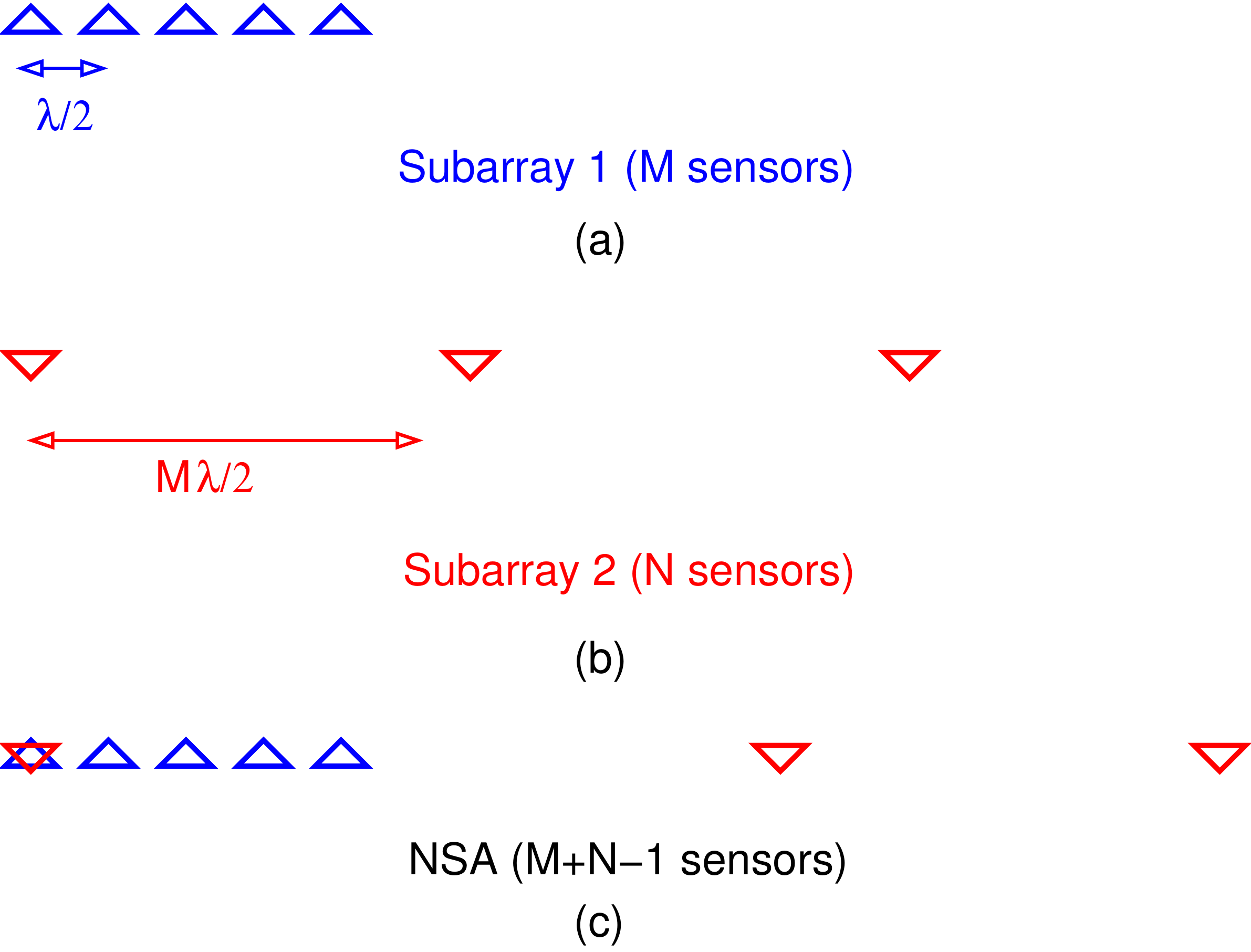}
\caption{(a) Subarray 1 (blue) of an NSA with $M=5$ and $N=3$ (2) Subarray 2 (red) of an NSA with  $M=5$ and $N=3$ (c) The resulting NSA.}
\label{nsasensors}
\end{figure}

A nested sensor array (NSA) is a sparse array that interleaves two ULAs, hereafter called Subarray 1, and Subarray 2. The Subarray 1 has $M$ sensors (Symbol \color{blue}\triblue\color{black} \hspace*{0.05cm} in Figure \ref{nsasensors}) and $\dfrac{\lambda}{2}$ intersensor spacing, and Subarray 2 has $N$ sensors (Symbol \color{red}\trired\color{black} \hspace*{0.05cm} in Figure \ref{nsasensors}) and $M\dfrac{\lambda}{2}$ intersensor spacing. The total number of sensors in an NSA is $M+N-1$ but it can achieve the resolution of a full ULA with $MN$ sensors. Figure \ref{nsasensors} depicts the formation of an NSA for $M=5,$ and $N=3.$  The  NSA has only $7$ sensors and it achieves the resolution of a full ULA with $15$ sensors.

An NSA processor eliminates the grating lobes by taking the product of the CBF beampatterns of the two subarrays. The Subarray 1 beampattern has only one major lobe at $u=0$ and it is the main lobe. The Subarray 1 has nulls at integer multiples of $2/M$.  The Subarray 2 beampattern  has the undersampling factor $M,$ and as a result, it has $M$ major lobes at integer multiples of $2/M.$ The major lobe at $u=0$ is the main lobe and the other $M-1$ major lobes are grating lobes due to undersampling. Since the grating lobes of the Subarray 2 are exactly at the nulls of the Subarray 1, taking the product of the two beampatterns removes the grating lobes.

\begin{figure}[!h]
\centering
\includegraphics[scale=0.4,trim = 0cm 4.5cm 0cm 4.5cm]{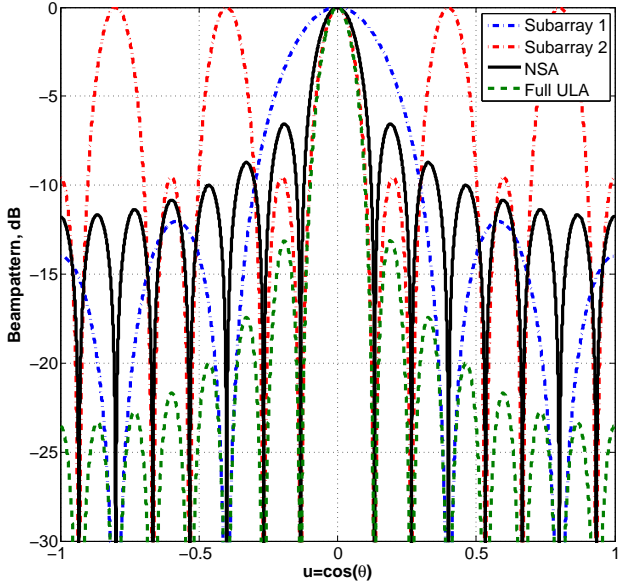}
\caption{Subarrays, nested and full ULA beampatterns}
\label{nestedbp}
\end{figure}

Figure \ref{nestedbp} illustrates the formation of an NSA beampattern for the arrays shown in Figure \ref{nsasensors}. The NSA has total $7$ sensors and the ULA has $15$ sensors. However, they have the equal main lobe width and therefore, have equal resolution. Comparing the NSA and the ULA beampatterns also shows that the NSA PSL height is much higher than the ULA PSL height. Extending the NSA subarrays like in CSA does not decrease the NSA PSL height to the level of the full ULA.

\subsubsection{Minimum Redundant Array}

Minimum redundant arrays is a class of sparse arrays designed to have the least redundancy in the coarray \cite[page 179]{VanTrees}. For a given number of sensors, an MRA attains the largest possible hole-free corrary while at the same time, minimizes the number of sensor pairs leading to the same spatial correlation lag. MRAs are the sparsest arrays among arrays that achieve hole-free coarrays. \cite[page 182]{VanTrees} lists the sensor locations for up to $17$ sensors. These sensor locations were obtained through exhaustive search routines and the analytical expressions for the sensor locations do not exist.

\subsubsection{Comparison of SCA with CSA, NSA, and MRA}

\begin{table}[h!]
\centering
\caption{\bf Comparison of the numbers of sensors relative to the full ULA with the equal resolution}\label{tab1}
  \begin{tabular*}{0.45\textwidth}{@{\extracolsep\fill}lcc@{\extracolsep\fill}}
    \hline
    \\[-0.5em]    
    \textbf{Array} & \textbf{Number of sensors} & \textbf{Matches ULA's}\\
         &  & \textbf{PSL height?}\\ \hline
             \\[-0.5em]
    ECSA & $\dfrac{2}{N}\left(\dfrac{13M+6}{13M+11}\right)$ & Yes \\
    \\[-0.5em]
    Basic CSA & $\dfrac{2}{N}$ & No \\
    \\[-0.5em]
    MCSA & $\dfrac{2}{N}$ & Yes \\    
        \\[-0.5em]
    NSA & $\dfrac{2}{N}$ & No \\
        \\[-0.5em]
    SCA & $\dfrac{2}{N}\left(\dfrac{PM+0.5Q-0.5}{PQM}\right)$ & Yes \\
    \\ \hline
        
  \end{tabular*}
  \end{table}

A sparse array achieves the resolution of a full ULA using fewer sensors. In addition to matching the resolution, some sparse arrays can match the PSL height of a full ULA. For a given resolution, the ratio of the number of sensors in the sparse array to the number of sensors in the full ULA provides a measure of the savings in sensors. The lower ratios indicate more savings in sensors. Table~\ref{tab1} lists the ratios of the total numbers of sensors of the sparse arrays to the total number of sensors of the full ULA with equal resolution (See Appendix \ref{app:totals}). The achievable ratio of the numbers of sensors is equal to $\dfrac{2}{N}$ for the basic CSA, MCSA, and NSA. For the ECSA, the achievable ratio is $\dfrac{2}{N}\left(\dfrac{13M+6}{13M+11}\right)$ which is less than $\dfrac{2}{N}$ for any $M$. For the SCA, the achievable ratio is $\dfrac{2}{N}\left(\dfrac{PM+0.5Q-0.5}{PQM}\right)$ which is also less than $\dfrac{2}{N}$ for any $M$. To realize the super-resolution feature embedded in the SCA, consider fixing the number of sensors to $L=32$. An ECSA with $M=2$ has $32$ sensors and it can match both the resolution and the PSL height of a $57$ sensor full ULA using only $32$ sensors. An MCSA with $M=8$ has $32$ sensors and it can match both the resolution and the PSL height of a $144$ sensor full ULA using only $32$ sensors. A basic CSA with $M=16$ has $32$ sensors and it can match the resolution of a $272$ sensor full ULA, but its PSL height is about $-5.5$ \rm{dB} which is much higher than the full ULA ($-13$ \rm{dB}). An NSA with $M=16$ and $N=17$ also has $32$ sensors and it can match the resolution of a $272$ sensor full ULA, but its PSL height is much higher than the full ULA. On the other hand, an SCA with $P=4$, $Q=9$, $M=3$ and $N=4$ has $32$ sensors and it can match both the resolution and the PSL of a $432$ sensor full ULA using only $32$ sensors. Hence, for a fixed number of sensors, the SCA achieves higher resolution than the other sparse arrays and at the same time, matches the PSL height of the full ULA.

\begin{figure}[!h]
\centering
\includegraphics[scale=0.4,trim = 0cm 6cm 0cm 4.5cm]{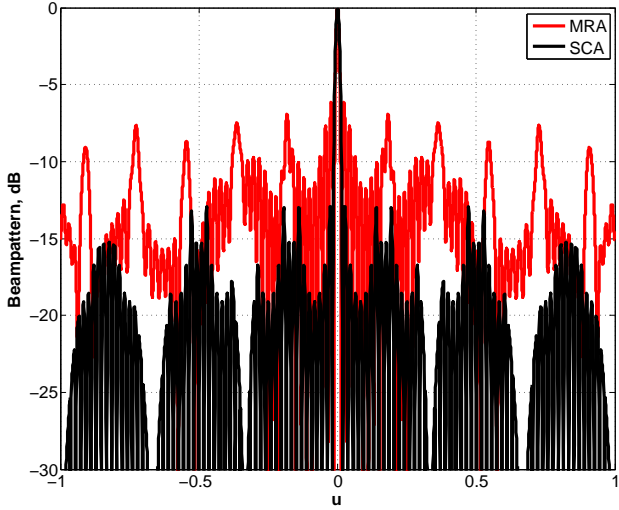}
\caption{Comparison of the MRA (red) and SCA (black) beampatterns. The two beampatterns exhibit equal main lobe width, however MRA's PSL height is much higher.}
\label{mrasca}
\end{figure}

Consider a minimum redundant array with number of sensors $L=17.$ This MRA can achieve the resolution of a $102$ sensor full array. An SCA with $P=3,$ $Q=6,$ $M=2$, and $N=3$ has $17$ sensors and it can achieve the resolution of a $102$ sensor full ULA. Hence, the SCA offers even more sparsity than the MRA. Moreover, the SCA exhibits better side lobe patterns than the MRA. The beampatterns of the SCA and MRA with $17$ sensors are shown in Figure \ref{mrasca}. The PSL height of the SCA is $-13$ \rm{db} whereas the PSL height of the MRA is $-6.1$ \rm{dB} as evident in Figure \ref{mrasca}.

\section{Estimation with Increased Degrees of Freedom}
\label{sec:results}

\begin{figure}[h]
\centering
\includegraphics[scale=0.43,trim = 0.5cm 9cm 0cm 4cm]{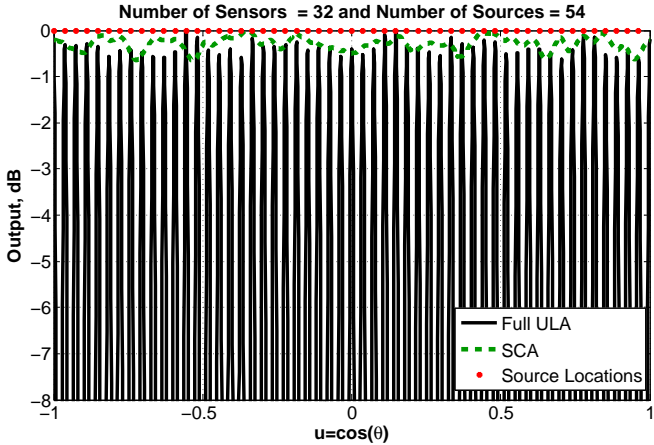}
\caption{DoA estimation comparison between an SCA (black solid) with parameters $M=3$, $N=4$, $P=5$, $Q=3$ and a ULA (green dashed) with the equal resolution. Number of Sensors = $32$, Number of Sources = $54$, Number of Snapshots = $100$ $SNR=0$ \rm{dB}}
\label{scaula}
\end{figure}

\begin{figure}[h]
\centering
\includegraphics[scale=0.43,trim = 0.5cm 9cm 0cm 4cm]{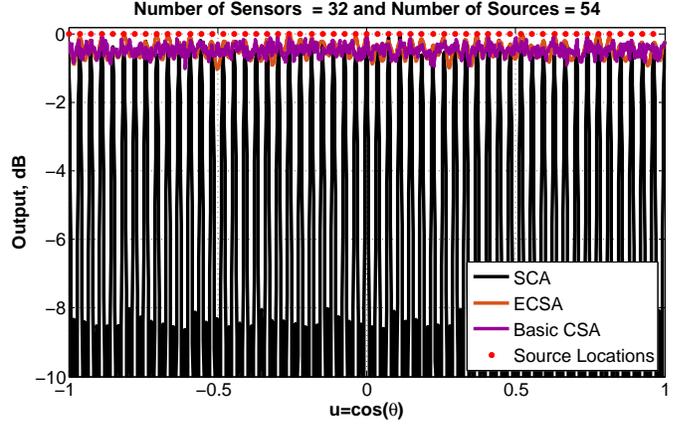}
\caption{DoA estimation comparison between an SCA (black solid) with parameters $M=3$, $N=4$, $P=5$, $Q=3$ and an ECSA (brown solid) with parameters $M_e=19$, $N_e=20$, $M=2$, $N=3$, $c=6.5$ and a basic CSA (purple solid) with parameters $M=16$, $N=17$.  Number of Sensors = $32$, Number of Sources = $54$, Number of Snapshots = $100$ $SNR=0$ \rm{dB}}
\label{scacsa}
\end{figure}

\begin{figure}[h]
\centering
\includegraphics[scale=0.43,trim = 0.5cm 9cm 0cm 4cm]{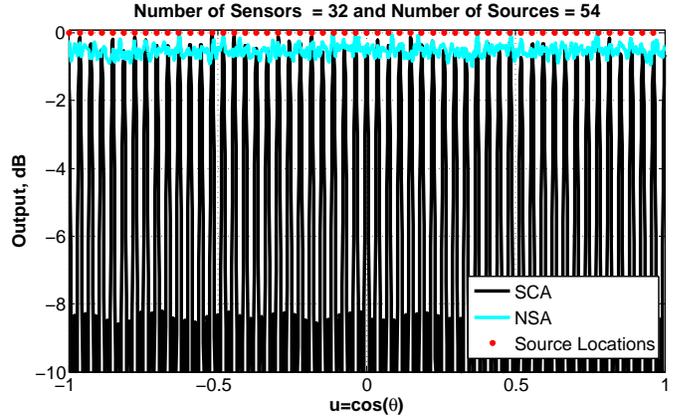}
\caption{DoA estimation comparison between an SCA with parameters $M=3$, $N=4$, $P=5$, $Q=3$ and an NSA (cyan solid) with parameters $M=10$, $N=23$. Number of Sensors = $32$, Number of Sources = $54$, Number of Snapshots = $100$ $SNR=0$ \rm{dB}}
\label{scansa}
\end{figure}

This section demonstrates SCA's ability to detect remarkably more sources than the number of sensors, and compares SCA's direction-of-arrival (DoA) estimation with the other spare arrays briefed in Section \ref{sec:comparison}. In all examples, the total number of sensors $L$ is $32$, the number of sources is $54$, the number of snapshots is $100$, and $SNR$ is $0$ \rm{dB}.

Figure \ref{scaula} compares DoA estimation by an SCA (black solid line) with parameters $M=3$, $N=4$, $P=5$, $Q=3$ with a full ULA (green dashed line). The $54$ source locations are marked by red circles in the figure. The SCA is able to detect all $54$ sources correctly. The full ULA cannot detect any source clearly in this example. Similarly, Figure \ref{scacsa} compares DoA estimation by an SCA with an ECSA ($M_e=19$, $N_e=20$, $M=2$, $N=3$, $c=6.5$), and a basic CSA ($M=16$, $N=17$). The SCA is able to identify all $54$ sources, whereas the CSAs fail to do so.  Finally, Figure \ref{scansa} compares DoA estimation by an SCA with an NSA with $M=10$ and $N=23.$ The SCA detects $54$ sources using only $32$ sensors, whereas the NSA does not. Hence, the degrees of freedom offered relative to the numbers of sensors is much higher in the SCA than in the other sparse arrays.

\section{Conclusion}

This paper introduced a novel sparse array called semi-coprime array, formed by interleaving three subarrays --- Subarray 1, Subarray 2, and Subarray 3. Subarray 1 and Subarray 2 have undersampling factors $QN$ and $QM$, where $M$ and $N$ are coprime integers and $Q$ is the number of sensors in Subarray 3. Min-processing the CBF beampatterns of the three subarrays removes the grating lobes from Subarray 1 and Subarray 2 that are at different locations, while the grating lobes that are at the same locations are suppressed by the nulls of Subarray 3. The resulting non-uniform sparse linear array offers more savings in the number of sensors than other sparse counterparts like coprime arrays, nested arrays, and minimum redundant arrays. Moreover, the SCA matches the PSL height of a full ULA more easily.

The super-resolution feature offered by the SCA can be exploited in applications that involve sensing and processing signals, for example signal estimation and detection. The examples presented in Section \ref{sec:results} shows that the SCA has the potential to offer much higher degrees of freedom relative to its number of sensors than possible with other existing sparse arrays, and this could lead to significant reduction in system cost and complexity.

\appendices

\section{Total number of sensors and resolution of a uniformly shaded ECSA}
\label{app:ecsa}

For an ECSA, the numbers of sensors in the Subarray 1 and Subarray 2  are $M_e=\lceil cN\rceil -1$ and $N_e=\lceil cN\rceil$, where $c$ is the smallest positive number that guarantees ECSA's PSL is less than or equal to $-13$ \rm{dB}. Since the subarrays share $\lceil c \rceil$ sensors, the total number of sensors in the array is
\begin{equation}
\nonumber
\begin{split}
L=&\lceil 2cN-1- c\rceil\\
=&\lceil 13N-1- 6.5\rceil \text{ since $c=6.5$ for uniform shading \cite{shadings}}\\
=&\lceil 13N-7.5\rceil\\
=&\lceil 13M+13-7.5\rceil \text{ since $N=M+1$.}\\
\end{split}
\end{equation}
Hence, the total number of sensors is $L=13M+6.$

Since the Subarray 1 has $M_e$ sensors and the intersensor spacing $N\dfrac{\lambda}{2},$ the main lobe width of its CBF beampattern is $\dfrac{4}{M_eN}$. Since the Subarray 2 has $M_e$ sensors and the intersensor spacing $M\dfrac{\lambda}{2},$ the main lobe width of its CBF beampattern is $\dfrac{4}{N_eM}$. The product of the two CBF beampatterns has the main lobe width
\begin{equation}
\nonumber
\begin{split}
MLW= &min\left( \dfrac{4}{M_eN},\dfrac{4}{N_eM}\right)\\
=&min\left( \dfrac{4}{(cN-1)(M+1)},\dfrac{4}{cNM}\right)\\
=&min\left( \dfrac{4}{cNM+cN-(M+1)},\dfrac{4}{cNM}\right)\\
=&min\left( \dfrac{4}{cNM+c(M+1)-(M+1)},\dfrac{4}{cNM}\right)\\
=&\dfrac{4}{cNM+c(M+1)-(M+1)}\text{ since $c>1$}.\\
\end{split}
\end{equation}

Hence, the main lobe width of the ECSA beampattern is equal to the main lobe width of the Subarray 1, $\dfrac{4}{M_eN}.$

\section{Total number of sensors and resolution of an MCSA}
\label{app:mcsa}

For an MCSA, when the numbers of sensors in the Subarray 1 and Subarray 2  are $M_e=2M$ and $N_e=2N$, the PSL height is close to $-13$ \rm{dB}. The total number of sensors in an MCSA with $N=M+1$, $M_e=2M$ and $N_e=2N$ is $L=2M+2N-2=4M.$ Each subarray has the main lobe width of $\dfrac{4}{2MN}$. Hence, the MCSA has the resolution equal to a full ULA with $2MN$ sensors.

\section{Ratios of the numbers of sensors in the sparse arrays to the full ULA}
\label{app:totals}

For an ECSA, the total number of sensors is $13M+6$ and it has the resolution of a full ULA with $M_eN$ sensors. Therefore, the ratio of the numbers of sensors in the ECSA to the full ULA is
\begin{equation}
\nonumber
\begin{split}
R=&\dfrac{13M+6}{M_eN}=\dfrac{2}{N}\dfrac{13M+6}{2M_e}\\
=& \dfrac{2}{N}\dfrac{13M+6}{(13N-2)} \text{ since $M_e=\lceil 6.5 N\rceil -1$}\\
=& \dfrac{2}{N}\dfrac{13M+6}{(13M+11)} \text{ since $N=M+1$}\\
\end{split}
\end{equation}

For a basic CSA, the total number of sensors is $M+N-1=2M$ and it has the resolution of a full ULA with $MN$ sensors. Therefore, the ratio of the numbers of sensors in the basic CSA to the full ULA is $R=\dfrac{2M}{MN}=\dfrac{2}{N}.$

For an MCSA, the total number of sensors is $2M+2N-2=4M$ and it has the resolution of a full ULA with $2MN$ sensors. Therefore, the ratio of the numbers of sensors in the MCSA to the full ULA is $R=\dfrac{4M}{2MN}=\dfrac{2}{N}.$

For an NSA, the total number of sensors is $M+N-1$ and it has the resolution of a full ULA with $MN$ sensors. Therefore, the ratio of the numbers of the sensors in the NSA to the full ULA is $R=\dfrac{M+N-1}{MN}.$ Assume $N=M-1$ for the ease of comparison with the other sparse arrays, which will result in $R=\dfrac{2}{N}.$

For an SCA, the total number of sensors is $PM+PN+Q-P-1$ and with $N=M+1$, the number of sensors is $2PM+Q-1$ and it has the resolution of a full ULA with $PQMN$ sensors. Therefore, the ratio of the numbers of sensors in the SCA to the full ULA is $\vspace*{0.1cm}R=\dfrac{2PM+Q-1}{PQMN}=\dfrac{2}{N}\dfrac{PM+0.5Q-0.5}{PQM}$

\section*{Acknowledgment}

The author would like to thank John R. Buck for helpful feedback on this research.

\ifCLASSOPTIONcaptionsoff
  \newpage
\fi

\bibliographystyle{IEEEtran}

\bibliography{IEEEabrv,SCAreferences}

\begin{thebibliography}{10}
\providecommand{\url}[1]{#1}
\csname url@samestyle\endcsname
\providecommand{\newblock}{\relax}
\providecommand{\bibinfo}[2]{#2}
\providecommand{\BIBentrySTDinterwordspacing}{\spaceskip=0pt\relax}
\providecommand{\BIBentryALTinterwordstretchfactor}{4}
\providecommand{\BIBentryALTinterwordspacing}{\spaceskip=\fontdimen2\font plus
\BIBentryALTinterwordstretchfactor\fontdimen3\font minus
  \fontdimen4\font\relax}
\providecommand{\BIBforeignlanguage}[2]{{%
\expandafter\ifx\csname l@#1\endcsname\relax
\typeout{** WARNING: IEEEtran.bst: No hyphenation pattern has been}%
\typeout{** loaded for the language `#1'. Using the pattern for}%
\typeout{** the default language instead.}%
\else
\language=\csname l@#1\endcsname
\fi
#2}}
\providecommand{\BIBdecl}{\relax}
\BIBdecl

\bibitem{Kpt}
D.~King, R.~Packard, and R.~Thomas, ``Unequally spaced broad-band antenna
  arrays,'' \emph{IRE Trans. Antennas Propag.}, vol. AP-8, pp. 380--385, July
  1960.

\bibitem{Unz}
H.~Unz, ``Linear arrays with arbitrarily distributed elements,'' \emph{IRE
  Trans. Antennas Propag.}, vol. AP-8, pp. 222--223, March 1960.

\bibitem{Ishimaru}
A.~Ishimaru, ``Theory of unequally spaced arrays,'' \emph{IRE Trans. Antennas
  Propag.}, vol. AP-10, pp. 691--702, November 1962.

\bibitem{IshimaruChen}
A.~Ishimaru and Y.~Chen, ``Thinning and broadbanding antenna arrays by unequal
  spacings,'' \emph{IRE Trans. Antennas Propag.}, vol. AP-13, pp. 34--42,
  January 1965.

\bibitem{Moffet}
A.~Moffet, ``Minimum-redundancy linear arrays,'' \emph{IRE Trans. Antennas
  Propag.}, vol.~16, pp. 172--175, March 1968.

\bibitem{steinberg}
B.~Steinberg, \emph{Principles of aperture and array system design: including
  random and adaptive arrays}, ser. A Wiley-Interscience publication.\hskip 1em
  plus 0.5em minus 0.4em\relax John Wiley and Sons Canada, Limited, 1976.

\bibitem{berman}
A.~Berman and C.~S. Clay, ``Theory of time averaged product arrays,'' \emph{The
  Journal of the Acoustical Society of America}, vol.~29, no.~7, 1957.

\bibitem{davis}
D.~Davies and C.~Ward, ``Low sidelobe patterns from thinned arrays using
  multiplicative processing,'' \emph{IEE Proceedings F Communications, Radar
  and Signal Processing}, vol. 127, no.~1, pp. 9--23, February 1980.

\bibitem{kefalas}
G.~Kefalas, ``An aperture distribution technique for product-array antennas,''
  \emph{IEEE Transactions on Antennas and Propagation}, vol.~16, no.~1, pp.
  125--125, January 1968.

\bibitem{mitra}
S.~Mitra, K.~Mondal, M.~Tchobanou, and G.~Dolecek, ``General polynomial
  factorization-based design of sparse periodic linear arrays,'' \emph{IEEE
  Transactions on Ultrasonics, Ferroelectrics, and Frequency Control}, vol.~57,
  no.~9, pp. 1952--1966, September 2010.

\bibitem{nested1}
P.~Pal and P.~Vaidyanathan, ``Nested arrays: A novel approach to array
  processing with enhanced degrees of freedom,'' \emph{IEEE Transactions on
  Signal Processing}, vol.~58, no.~8, pp. 4167 --4181, {A}ugust 2010.

\bibitem{nested2}
------, ``Nested arrays in two dimensions, part {I}: Geometrical
  considerations,'' \emph{IEEE Transactions on Signal Processing}, vol.~60,
  no.~9, pp. 4694 --4705, {S}eptember 2012.

\bibitem{VandP1}
P.~Vaidyanathan and P.~Pal, ``Sparse sensing with co-prime samplers and
  arrays,'' \emph{IEEE Transactions on Signal Processing}, vol.~59, no.~2, pp.
  573--586, {F}ebruary 2011.

\bibitem{VandP2}
------, ``Theory of sparse coprime sensing in multiple dimensions,'' \emph{IEEE
  Transactions on Signal Processing}, vol.~59, no.~8, pp. 3592--3608, {A}ugust
  2011.

\bibitem{QinZhangAmin}
S.~Qin, Y.~Zhang, and M.~Amin, ``Generalized coprime array configurations for
  direction-of-arrival estimation,'' \emph{IEEE Transactions on Signal
  Processing}, vol.~63, pp. 1377 -- 1390, {M}arch 2015.

\bibitem{KBW}
K.~Adhikari, J.~Buck, and K.~Wage, ``Beamforming with extended co-prime sensor
  arrays,'' \emph{2013 IEEE International Conference on Acoustics, Speech and
  Signal Processing (ICASSP)}, pp. 4183--4186, May 2013.

\bibitem{shadings}
\BIBentryALTinterwordspacing
------, ``Extending coprime sensor arrays to achieve the peak side lobe height
  of a full uniform linear array,'' \emph{EURASIP Journal on Advances in Signal
  Processing}, vol. 2014, no.~1, p. 148, {S}ep 2014. [Online]. Available:
  \url{https://doi.org/10.1186/1687-6180-2014-148}
\BIBentrySTDinterwordspacing

\bibitem{liubuck1}
Y.~Liu and J.~Buck, ``Detecting gaussian signals in the presence of interferers
  using the coprime sensor arrays with the min processor,'' \emph{2015 49th
  Asilomar Conference on Signals, Systems and Computers}, pp. 370--374, Nov
  2015.

\bibitem{liubuck2}
------, ``Super-resolution doa estimation using a coprime sensor array with the
  min processor,'' \emph{2016 50th Asilomar Conference on Signals, Systems and
  Computers}, pp. 944--948, Nov 2016.

\bibitem{liubuck3}
------, ``Spatial spectral estimation using a coprime sensor array with the min
  processor,'' \emph{2016 IEEE Sensor Array and Multichannel Signal Processing
  Workshop (SAM)}, pp. 1--5, July 2016.

\bibitem{VanTrees}
H.~V. Trees, \emph{Optimum Array Processing (Detection, Estimation and
  Modulation Theory, Part IV)}.\hskip 1em plus 0.5em minus 0.4em\relax John
  Wiley and Sons, Inc., New York, 2002.

\end{thebibliography}

\end{document}